\documentstyle[12pt,twoside,fleqn,espcrc1,psfig]{article}

\newcommand{\AmS}{{\protect\the\textfont2
  A\kern-.1667em\lower.5ex\hbox{M}\kern-.125emS}}
\title{Applicability of the Hauser-Feshbach approach for the determination
of astrophysical reaction rates}

\author{T. Rauscher\address{Institut f\"ur Physik, 
        Universit\"at Basel, \\ 
        Klingelbergstr.\ 82, CH-4056 Basel, Switzerland}%
        \thanks{APART fellow of the Austrian Academy of Sciences},
F.-K. Thielemann$^{\mathrm{a}}$, and
K.-L. Kratz\address{Institut f\"ur Kernchemie, Universit\"at Mainz,\\
Fritz-Strassmann-Weg 2, D-55099 Mainz, Germany}}

\begin{document}
% typeset front matter
\maketitle

\begin{abstract}
Nuclear Astrophysics requires the knowledge of reaction rates over a wide
range of nuclei and temperatures. In recent calculations the nuclear
level density -- as an important ingredient to the statistical model
(Hauser-Feshbach) --
has shown the highest uncertainties. In a back-shifted fermi-Gas
formalism utilizing an energy-dependent level density parameter and
employing microscopic corrections from a recent FRDM mass formula, we obtain
a highly improved fit to experimental level densities. The resulting
level density is used for determining criteria for the applicability of the
statistical model.
\end{abstract}

\section{INTRODUCTION}

The field of Nuclear Astrophysics has to provide nuclear reaction rates
suited for a wide range of astrophysical applications. Therefore, there is
not only need for rates involving all possible (stable and unstable) nuclei
across the nuclear chart but also for temperatures ranging from 
$0<T_9<10$ . For the majority of reactions, the statistical model
(Hauser-Feshbach, SM) will be suited for the calculation and prediction of the
rates. However, at low level densities the SM 
is not applicable anymore and single resonances and direct capture
contributions have to be taken into account. 

For the user of such rates it would be convenient to have a simple means to
determine the validity of the SM approach, showing which rates need special
attention and probably further experimental investigation. In this work, we
provide a ``map'' for the applicability of the SM depending on the
interacting nuclei and the temperature.

\section{THE NUCLEAR LEVEL DENSITY}

Considerable effort has been put into the improvement of the input for
the SM calculations (e.g.~\cite{cowan}). However,
the nuclear level density still has shown the largest uncertainties among
the properties entering the SM. For
calculating the
level densities in the given context one does not only have to find reliable
methods, but also computationally feasible ones. In dealing with
thousands of nuclei one has to resort to simple models in order to
minimize computer time. 

Such a simple model is the back-shifted Fermi-gas
description~\cite{cowan,bethe}, recently improved by introducing an
energy-dependent level density parameter $a$~\cite{igna,reis,ilji}. 
More sophisticated Monte Carlo shell 
model calculations~\cite{karli} have shown excellent agreement with this 
phenomenological approach and justified the application of the Fermi-gas
description at and above the neutron separation energy. Assuming equally
distributed even and odd parities, one obtains the following form:
\begin{equation}
\label{levden}
\rho(U,J,\pi)={1 \over 2} f(U,J) \rho(U)\quad,
\end{equation}
with
\begin{eqnarray}
\rho(U)={1 \over \sqrt{2\pi} \sigma}{\sqrt{\pi} \over
12a^{1/4}}{\exp(2\sqrt{aU}) \over U^{5/4}}\ ,\qquad
f(U,J)={2J+1 \over 2\sigma^2} \exp\left({-J(J+1) \over
2\sigma^2}\right) \\
\sigma^2={\Theta_{\mathrm{rigid}} \over \hbar^2} \sqrt{U \over a}\ ,\qquad
\Theta_{\mathrm{rigid}}={2 \over 5}m_{\mathrm{u}}AR^2\ ,\qquad
U=E-\delta\quad. \nonumber
\end{eqnarray}
An improved approach has to consider the energy dependence of the microscopic
effects which are known to vanish at high excitation energies~\cite{ilji}, 
i.e.\ the thermal damping of microscopic effects.
The level density parameter $a$ is then described by~\cite{igna}
\begin{equation}
\label{endepa}
a(U,Z,N)=\tilde{a}(A)\left[1+C(Z,N){f(U) \over
U}\right]\quad,
\end{equation}
where
\begin{equation}
\tilde{a}(A)=\alpha A+\beta A^{2/3}
\end{equation}
and
\begin{equation}
f(U)=1-\exp(-\gamma U)\quad.
\end{equation}
The shape of the function $f(U)$ was found by approximation of
numerical microscopic calculations based on the shell
model. Thus, one is left with three open parameters,
namely $\alpha$, $\beta$, and $\gamma$. The values of these parameters
are determined by a fit to experimental s-wave neutron resonance spacing
at the neutron separation energy~\cite{ilji}. The values
$\alpha=0.1336$, $\beta=-0.06712$, $\gamma=0.04862$ result in a highly
improved fit with an averaged global deviation of 1.5~\cite{tomturin,tom},
when taking the microscopic corrections $C(Z,N)$ from the latest FRDM
mass formula~\cite{moeller} and consistently computing the backshift 
$\delta(Z,N)$=1/2\{$\Delta_{\mathrm{n}}(Z,N)+\Delta_{\mathrm{p}}(Z,N)$\} with
the neutron and proton pairing gaps $\Delta_{\mathrm{n,p}}$ from the same
source.

\section{APPLICABILITY OF THE STATISTICAL MODEL}

Having found a suitable method to calculate level densities one can
apply it to determine the range of validity of the SM.
It is often colloquially termed that the SM is
only applicable for intermediate and heavy nuclei. However, the only necessary
condition for its application is a large number of resonances
at the appropriate bombarding energies, so that the cross section can be
described by an average over resonances.

\begin{figure}
%\vspace*{18cm}
\psfig{file=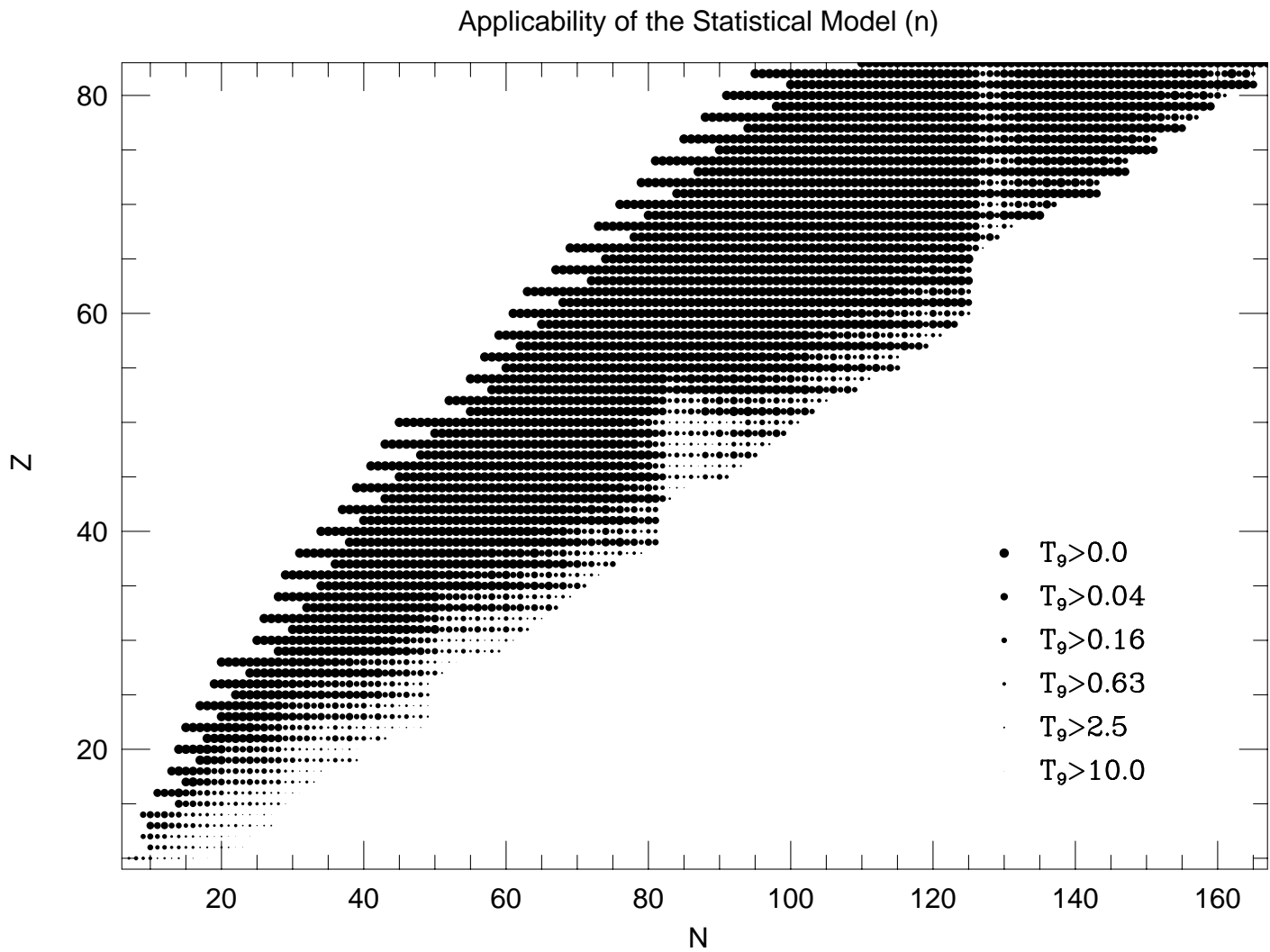,height=21.9cm,rheight=20.8cm}
\caption{\label{nfig}Temperatures (in T$_9$) for which the statistical
model can be used. Plotted is the compound nucleus of the neutron-induced
reaction.}
\end{figure}
The nuclear reaction rate per particle pair at a given stellar temperature
$T$ is determined by folding the reaction cross section with the
Maxwell-Boltzmann (MB) velocity distribution of the projectiles~\cite{fowler}
\begin{equation}
\left< \sigma v \right>=\left( \frac{8}{\pi \mu} \right) ^{1/2}
\frac{1}{\left( kT \right) ^{3/2}}
\int_0^{\infty} \sigma(E) E \exp \left( -\frac{E}{kT} \right) dE \quad.
\end{equation}
An effective energy window is then found around the peak
of the integrand at $E_0$.
For charged particles this is the so-called Gamow peak at 
$E_0=E_{\mathrm{G}}^{1/3}(kT/2)^{2/3}$ (with
the Gamow energy $E_{\mathrm{G}}$). For s-wave neutrons the effective
peak coincides with the peak of the MB distribution at
$E_0=kT$ (close to the neutron separation energy), for higher partial waves 
the energy window is shifted to slightly higher energies (similarily to the
Gamow peak) due to the centrifugal barrier~\cite{wagoner}. The effective
energy window for a given nucleus and temperature has to contain
sufficiently many resonances in order to make it possible to solve the integral
with the assumption of an average level density instead of calculating the
exact sum over the individual levels. Numerical test
calculations~\cite{tom} have shown that an average number of 5--10 
contributing resonances is sufficient. Choosing a lower limit for the number
of resonances, determining the width and location of the effective energy
window at a given temperature and using the above level density description
%(Eq.~\ref{levden}) 
to calculate the number of resonances in this window, 
we derive a lower limit for the temperature at which the SM can still be used. 
Those temperature limits are shown in Fig.~\ref{nfig} for neutron-induced
reactions. It should be noted that the derived temperatures will not change
considerably even if changing the required level number within a factor
of two, because of the exponential dependence of the level density on the
excitation energy.

\section{SUMMARY}

Making use of an improved level density description we presented a method
to determine the applicability of the SM, also providing clues on
which reactions might be of special in\-ter\-est for experimental 
investigations.
In principle, the method can be used for any pro\-jec\-ti\-les 
and also with different
incident energy distributions (e.g.\ for experimental beams).

\end{document}